%% file: iulp-X.tex
\documentclass{llncs}

\usepackage[utf8]{inputenc}
\usepackage{amsmath,amssymb}
\usepackage{times,helvet,courier}
\usepackage{graphicx}
\usepackage{url}

\usepackage{listings}
\input{macro}

\title{Interactive Answer Set Programming\\ --- Preliminary Report ---}

\author{%
  Martin Gebser\inst{1,3}
  \and
  Philipp Obermeier\inst{3}
  \and
  Torsten Schaub\inst{2,3}\thanks{Affiliated with the
                              School of Computing Science at
                              Simon Fraser University,
                              Burnaby, Canada,
                              and the
                              Institute for Integrated and Intelligent Systems
                              at
                              Griffith University,
                              Brisbane, Australia.}
  }

\institute{Aalto University, HIIT \and INRIA Rennes \and University of Potsdam}

\begin{document}

\maketitle

\input{abstract}

\input{introduction}
\input{approach}

\input{system}

\input{discussion}

\input{acknowledgments}


\input{bbl}

\end{document}

%% file: macro.tex
\newcommand{\naf}[1]{\ensuremath{{\sim\!{#1}}}}
\newcommand{\naff}[1]{\ensuremath{{\sim{#1}}}}

\newcommand{\head}[1]{\ensuremath{\mathit{head}(#1)}}
\newcommand{\Head}[1]{\ensuremath{\mathit{head}(#1)}}
\newcommand{\body}[1]{\ensuremath{\mathit{body}(#1)}}

\newcommand{\atom}[1]{\ensuremath{\mathit{atom}(#1)}}

\newcommand{\poslits}[1]{\ensuremath{{#1}^+}}
\newcommand{\neglits}[1]{\ensuremath{{#1}^-}}

\newcommand{\pbody}[1]{\poslits{\body{#1}}}
\newcommand{\nbody}[1]{\neglits{\body{#1}}}

\newcommand{\PRG}{\ensuremath{P}}
\newcommand{\QRG}{\ensuremath{Q}}
\newcommand{\RRG}{\ensuremath{R}}

\newcommand{\AS}{\ensuremath{\mathit{AS}}}

\newcommand{\sysfont}{\textit}
\newcommand{\aspic}{\sysfont{aspic}}

\newcommand{\clingo}{\sysfont{clingo}}

\newcommand{\gringo}{\sysfont{gringo}}

\newcommand{\oclingo}{\sysfont{oclingo}}

\newcommand{\quontroller}{\sysfont{quontroller}}

\newcommand{\query}{\ensuremath{\mathit{query}}} 

\newcommand{\assume}{\ensuremath{\mathit{assume}}} 
\newcommand{\cancel}{\ensuremath{\mathit{cancel}}} 
\newcommand{\assert}{\ensuremath{\mathit{assert}}} 
\newcommand{\retract}{\ensuremath{\mathit{retract}}} 
\newcommand{\open}{\ensuremath{\mathit{open}}} 
\newcommand{\define}{\ensuremath{\mathit{define}}} 
\newcommand{\external}{\ensuremath{\mathit{external}}} 
\newcommand{\release}{\ensuremath{\mathit{release}}} 
\newcommand{\filter}{\ensuremath{\mathit{\nu}}}

\newcommand{\Q}{\ensuremath{Q}}

\newcommand{\etmcuap}{\ensuremath{\mu}}

\newcommand{\CONFINEOP}{\ensuremath{C}}
\newcommand{\CONFINE}[2]{\ensuremath{\CONFINEOP_{#1}(#2)}}


%% file: abstract.tex
\begin{abstract}
  Traditional Answer Set Programming (ASP) rests upon one-shot solving.
  A logic program is fed into an ASP system and its stable models are computed.
  The high practical relevance of dynamic applications led to the development of multi-shot solving systems.
  An operative system solves continuously changing logic programs.
  Although this was primarily aiming at dynamic applications in assisted living, robotics, or stream reasoning, where solvers interact with an environment,
  it also opened up the opportunity of interactive ASP, where a solver interacts with a user.
  We begin with a formal characterization of interactive ASP in terms of states and operations on them.
  In turn, we describe the interactive ASP shell \textit{aspic} along with its basic functionalities.
\end{abstract}

%% file: introduction.tex
\section{Introduction}\label{sec:introduction}

Traditional logic programming \cite{clomel81,lloyd87} is based upon query answering.
Unlike this,
logic programming under the stable model semantics \cite{gellif88b} is commonly implemented by model generation based systems,
viz.\ answer set solvers~\cite{gekakasc12a}.
Although the latter also allows for checking whether a (ground) query is entailed by some or all stable models, respectively,
there is no easy way to explore a domain at hand by posing consecutive queries without relaunching the solver.
The same applies to the interactive addition and/or deletion of temporary program parts that come in handy
during theory exploration, for instance, when dealing with hypotheses.

An exemplary area where such exploration capacities would be of great benefit is bio-informatics
(cf.\ \cite{bachtrtrjobe04a,erdtur08a,gescthve10a,geguivscsithve10a,rawhki10a,viguedthgrsasi12a}).
Here, we usually encounter problems with large amounts of data, 
resulting in runs having substantial grounding and solving times.
Furthermore, 
problems are often under-constrained, thus yielding numerous alternative
solutions.
In such a setting, it would be highly beneficial to explore a domain via successive queries and/or
under certain hypotheses.
For instance,
for determining nutritional requirements for sustaining maintenance or growth of an organism,
it is important to indicate seed compounds needed for the synthesis of 
other compounds~\cite{schthi09a}.
Now, rather than continuously analyzing several thousand stable models (or their intersection or
union), a biologist may rather perform interactive ``in-silico'' experiments by temporarily adding 
compounds and subsequently exploring the resulting models by posing successive queries.

We address this shortcoming and show how the multi-shot solving capacities of \clingo~4~\cite{gekakasc14b} 
can be harnessed to provide query answering and theory exploration functionalities.
In fact, multi-shot ASP was conceived for incorporating online information into operative ASP solving processes.
Although this technology was originally devised for dealing with data streams in dynamic environments,
like assisted living or cognitive robotics,
it can likewise be used to incorporate facts, rules, or queries provided by a user.
As a result,
we present the semantics, design, and implementation of a system for interactive query answering and theory exploration with ASP.
Our ASP shell \aspic\ is based on the the answer set solver \clingo~4
and is implemented as a dedicated front-end.

Our approach along with the resulting system greatly supersede the one of the earlier prototype \quontroller~\cite{geobsc13a},
which itself was based on the archetypal reactive solver \oclingo.
The \clingo~4 series goes beyond its predecessors by providing control capacities
that can be used either through scripting inlets in ASP encodings or corresponding APIs
in Lua and Python.%
\footnote{%
Details can be found in~\cite{gekakasc14b}.
Similarly, we refer the reader to the literature for introductions to ASP~\cite{gelkah14a,gekakasc12a} and the input language of \clingo~4~\cite{PotasscoUserGuide}.}
Given this,
\aspic\ does not only provide a much more general framework than \quontroller,
but is moreover based on firm semantic underpinnings that we develop in Section~\ref{sec:approach}.
In Section~\ref{sec:system}, we illustrate the functionality and implementation of \aspic,
and Section~\ref{sec:discussion} concludes the paper.


%% file: approach.tex
\section{Operational Semantics}\label{sec:approach}

This section defines an operational semantics in terms of system states and operations modifying such states.
We confine ourselves to a propositional setting but discuss query answering with variables
in Section~\ref{sec:non-ground}.
To begin with, we introduce the major ingredients of our semantics.

We consider logic programs and assignments over an alphabet $\mathcal{A}$ of ground atoms.
A rule $r$ is 
of the form
\(
h \leftarrow \ell_1,\dots,\ell_n
\)
with head
$h=a$, $h=\{a\}$, or $h=\bot$, where $a\in\mathcal{A}$ and $\bot$ is a constant denoting falsity,
and body $\ell_1,\dots,\ell_n$ such that $\ell_i=a_i$ or $\ell_i=\naf{a_i}$
for $1\leq i\leq n$, where $a_i\in \mathcal{A}$ and $\sim$ stands for default negation.
%
We denote the head atom~$a$ or~$\bot$, respectively, by \head{r},
and $\body{r}=\{\ell_1,\dots,\ell_n\}$ is the set of body literals of~$r$.
Given a literal $\ell$, we define $\overline{\ell}$ as $\naf{a}$, if $\ell=a$, and as $a$, if $\ell=\naf{a}$, for $a\in\mathcal{A}$.
For a set $L$ of literals,
let $\poslits{L}=L\cap\mathcal{A}$ and $\neglits{L}=\{\overline{\ell}\mid\ell\in L\setminus\mathcal{A}\}$
denote the sets of atoms occurring positively or negatively, respectively, in~$L$.
A logic program~\PRG\ is a finite set of rules.
The set of head atoms of~\PRG\ is
$\head{\PRG}=\{\head{r}\in\mathcal{A} \mid r\in\PRG\}$;
note that $\bot\notin\head{\PRG}$ even when $\head{r}=\bot$ for some $r\in\PRG$,
and below we skip~$\bot$ when writing rules without head atom.
Moreover, we use \atom{\phi} to refer to all atoms occurring in $\phi$, no matter whether $\phi$ is a program or any other type of logical expression.
Finally, we let $\AS(\PRG)$ stand for all stable models of
~\PRG.

An assignment $i$ over a set $A\subseteq\mathcal{A}$ of atoms is a function $i: A\rightarrow\{t,f,u\}$.
We often refer to an assignment $i$ in terms of three characteristic functions:
$i^t=\{a\in\mathcal{A}\mid i(a)=t\}$, 
$i^f=\{a\in\mathcal{A}\mid i(a)=f\}$, and 
$i^u=\{a\in\mathcal{A}\mid i(a)=u\}$.

Given this,
a \emph{system state} over $\mathcal{A}$ is defined as a quadruple
\[
(\RRG,I,i,j)
\]
where
\begin{itemize}
\item \RRG\ is a ground program over $\mathcal{A}$,
\item $I \subseteq \mathcal{A} \setminus \Head{R}$ is a set of input atoms,
\item $i$ is an assignment over $I$, and
\item $j$ is an assignment over $\mathcal{A}$. 
\end{itemize}
A system state $(\RRG,I,i,j)$ induces the following logic program
\begin{align*}
  \PRG(\RRG,I,i,j)=\RRG &\cup \{a\leftarrow{}\mid a\in i^t\}       \cup\{\{a\}\leftarrow{}\mid a \in i^u\}\\
                        &\cup \{{}\leftarrow\naf{a}\mid a\in j^t\} \cup \{{}\leftarrow a\mid a\in j^f\}
\end{align*}
Note the different effects of assignments $i$ and $j$.
Changing truth values in $i$ amounts to data manipulation,
while changing the ones in $j$ boils down to a data filter.
In fact,
atoms true in $i$ are exempt from the unfoundedness criterion,
``undefined''\footnote{Strictly speaking, such atoms are not undefined but assigned value $u$.} atoms give rise to two alternatives,
and false atoms are left to the semantics and thus set to false.
Unlike this, assignment $j$ acts as a mere filter on the set  of stable models;
in particular, atoms true in $j$ are subject to the unfoundedness criterion, while
an atom ``undefined'' in $j$ imposes no constraint.
Accordingly, $i$ and $j$ have also a different default behavior.
While $i$ defaults to false, $j$ defaults to ``undefined'',
and we thus leave $a\mapsto f$ or $a\mapsto u$, respectively,
implicit when specifying assignments $i$ and $j$.

\subsection{State Changing Operations}\label{sec:grd_scops}

We now start by fixing the operational semantics of our approach in terms of state changing operations.
A state changing operation is a function mapping a system state to another, 
depending on an extra input parameter.
%

\subsubsection{Assumption and Cancellation.}

Operator \assume\ takes a ground literal $\ell$ and transforms a system state as follows
\[
\assume: \langle\ell,(R,I,i,j_1)\rangle \mapsto (R,I,i,j_2)
\]
where
\begin{itemize}
\item $j_2^t=j_1^t\cup\{ \ell \}$ if $ \ell \in\mathcal{A}$, and $j_2^t=j_1^t\setminus\{\overline{\ell}\}$ otherwise
\item $j_2^f=j_1^f\cup\{\overline{\ell}\}$ if $\overline{\ell}\in\mathcal{A}$, and $j_2^f=j_1^f\setminus\{ \ell \}$ otherwise
\item $j_2^u=\mathcal{A}\setminus (j_2^t\cup j_2^f)$
\end{itemize}

For example,
\(
\assume(a,(\{\{a\}\leftarrow{}\},\emptyset,\emptyset,\emptyset))
\)
is associated with program
\[
\PRG(\{\{a\}\leftarrow{}\},\emptyset,\emptyset,\{a\mapsto t\})
=
\{\{a\}\leftarrow{}\}\cup\{{}\leftarrow \naf{a}\}
\]
having the stable model $\{a\}$ only.
Unlike this,
\(
\assume(a,(\emptyset,\emptyset,\emptyset,\emptyset))
\)
induces program
\(
\PRG(\emptyset,\emptyset,\emptyset,\{a\mapsto t\})
=
\{{}\leftarrow \naf{a}\}
\),
which has no stable model.

Operator $\cancel$ takes a ground literal $\ell$ and transforms a system state as follows
\[
\cancel: \langle\ell,(R,I,i,j_1)\rangle\mapsto (R,I,i,j_2)
\]
where
\begin{itemize}
\item $j_2^t=j_1^t\setminus\{          \ell \}$ 
\item $j_2^f=j_1^f\setminus\{\overline{\ell}\}$ 
\item $j_2^u=\mathcal{A}\setminus (j_2^t\cup j_2^f)$
\end{itemize}

For example,
\(
\cancel(a,\assume(a,(\{\{a\}\leftarrow{}\},\emptyset,\emptyset,\emptyset)))
\)
yields the two stable models of $\{\{a\}\leftarrow{}\}$.
More generally, we observe the following interrelationships 
between the above operations.
For a system state $S=(\RRG,I,i,j)$ and a literal $\ell$:
\begin{align*}
  \cancel(\ell,\assume(\ell,S)) &= S\text{ if }\ell\notin  j^t\cup j^f \text{ and }\overline{\ell}\notin  j^t\cup j^f \\
  \cancel(\ell,\cancel(\ell,S)) &= \cancel(\ell,S)\\
  \assume(\ell,\cancel(\ell,S)) &= \assume(\ell,S)\\
  \assume(\ell,\assume(\ell,S)) &= \assume(\ell,S)
\end{align*}

\subsubsection{Assertion, Opening, and Retraction.}

Operator $\assert$ takes a ground atom $a$ and transforms a system state as follows
\[
\assert: \langle a,(R,I,i_1,j)\rangle \mapsto (R,I,i_2,j)
\]
where
\begin{itemize}
\item $i_2^t=i_1^t\cup\{a\}$      if $a\in I$, and $i_2^t=i_1^t$ otherwise
\item $i_2^u=i_1^u\setminus\{a\}$ 
\item $i_2^f=I\setminus (i_2^t\cup i_2^u)$
\end{itemize}
Note that, unlike with \assume, the atom of an asserted literal must belong to the input atoms.

For example,
\(
\assert(a,(\emptyset,
           \{a\},
                 \emptyset,\emptyset))
\)
is associated with program
\[
\PRG(\emptyset,
     \{a\},\{a\mapsto t\},\emptyset)
=
\{a\leftarrow{}\}
\]
having the only stable model $\{a\}$.

Operator $\open$ takes a ground atom $a$ and transforms a system state as follows
\[
\open: \langle a,(R,I,i_1,j)\rangle \mapsto (R,I,i_2,j)
\]
where
\begin{itemize}
\item $i_2^t=i_1^t\setminus\{a\}$ 
\item $i_2^u=i_1^u\cup\{a\}$ if $a\in I$, and $i_2^u=i_1^u$ otherwise
\item $i_2^f=I\setminus (i_2^t\cup i_2^u)$
\end{itemize}

Continuing the above example,
\(
\open(a,
\assert(a,(\emptyset,
           \{a\},
                 \emptyset,\emptyset))
)
=
\open(a,\linebreak[1](\emptyset,\linebreak[1]
         \{a\},\linebreak[1]\{a\mapsto t\},\linebreak[1]\emptyset)
)
\)
leads to
\[
\PRG(\emptyset,
     \{a\},\{a\mapsto u\},\emptyset)
=
\{\{a\}\leftarrow{}\}
\]
having the empty stable model in addition to $\{a\}$.

Operator $\retract$ takes a ground atom $a$ and transforms a system state as follows
\[
\retract: \langle a,(R,I,i_1,j)\rangle \mapsto (R,I,i_2,j)
\]
where
\begin{itemize}
\item $i_2^t=i_1^t\setminus\{a\}$ 
\item $i_2^u=i_1^u\setminus\{a\}$ 
\item $i_2^f=I\setminus (i_2^t\cup i_2^u)$
\end{itemize}

For example,
\(
\retract(a,
(\emptyset,
 \{a\},\{a\mapsto t\},\emptyset)
)
\)
yields
\(
\PRG(\emptyset,\linebreak[1]
     \{a\},\linebreak[1]
                        \emptyset,\linebreak[1]\emptyset)
=\nolinebreak
\emptyset
\),
whose stable model is empty as well.
More generally, we observe the following interrelationships between the above operations.
For a system state $S=(\RRG,I,i,j)$ and an atom~$a$:%
\begin{align*}
  \assert(a,S) &= S\text{ if }a\notin I\\
  \open(a,S)   &= S\text{ if }a\notin I\\
  \retract(a,\assert(a,S)) &= S\text{ if }a\notin  i^t\cup i^u \\
  \retract(a,\open(a,S))   &= S\text{ if }a\notin  i^t\cup i^u \\
  o(a,o'(a,S)) &= o(a,S)\text{ for }o,o'\in\{\assert,\open,\retract\}
\end{align*}

\subsubsection{Definition, External, and Release.}

Operator $\define$ takes a set $R$ of ground rules  
and transforms a system state as follows
\[
\define: \langle R,(R_1,I_1,i,j)\rangle \mapsto (R_2,I_2,i,j)
\]
where
\begin{itemize}
\item $R_2=\CONFINE{I_1}{R_1\cup R}$ if $R_1$ and $R$ are compositional, 
  and $R_2=R_1$ otherwise
\item $I_2=I_1\setminus\Head{R_2}$
\end{itemize}
Function \CONFINEOP\ confines a program to the atom base of the current system state.
More precisely,
a program \PRG\ and a set $I$ of input atoms induce the atom base $B=I\cup\head{\PRG}$
with which we define \CONFINE{I}{\PRG} as:
\[
\{\head{r}\leftarrow\pbody{r}\cup\{\naf{a}\mid a\in\nbody{r}\cap B\}\mid r\in \PRG,\pbody{r}\subseteq B\}
\]
This restricts the rules in \PRG\ to all atoms available as heads in \PRG\ or as input atoms in $I$.

Given that \clingo's program composition follows the one of module theory~\cite{oikjan06a},
the same criteria apply to the implementation of \define\ in \aspic:
Two programs \PRG\ and \QRG\ are \emph{compositional}, if
\begin{itemize}
\item $\Head{\PRG} \cap \Head{\QRG} = \emptyset$ and
\item for every strongly connected component $C$ in the positive dependency graph of $\PRG\cup\QRG$, 
  we have $\Head{\PRG}\cap C=\emptyset$ or $\Head{\QRG}\cap C=\emptyset$.
\end{itemize}
In words,
atoms must not be redefined and no mutual positive recursion is permissible in the joint program.
More liberal definitions are possible but not supported by the current framework of \aspic.

As an example, consider
\(
\define(\{a\leftarrow\naf{b}\},(\emptyset,\emptyset,\emptyset,\emptyset))
\).
The state and added program induce atom base $\{a\}$.
With it, the resulting state is captured by the simplified program
\(
\{a\leftarrow{}\}
\),
having a single stable model containing $a$.
Adding afterwards $b$ via
\(
\define(\{b\leftarrow{}\},\define(\{a\leftarrow\naf{b}\},(\emptyset,\emptyset,\emptyset,\emptyset))
\)
results in program
\(
\{a\leftarrow{},b\leftarrow{}\}
\)
along with a stable model containing $a$ and $b$. 

For another example, consider the operation
\(
\define(\{a\leftarrow{b},b\leftarrow{a}\},(\emptyset,\emptyset,\emptyset,\emptyset))
\).
This state results in an empty stable model, 
whereas the operator composition
\(
\define(\{a\leftarrow{b}\},\define(\{b\leftarrow{a}\},(\emptyset,\{a\},\emptyset,\emptyset)))
\)
is noneffective because a positive cycle would be obtained in the joined program.

Similarly, operation
\(
\define(\{a\leftarrow{b},a\leftarrow\naf{c}\},(\emptyset,\{b,c\},\emptyset,\emptyset))
\)
is well-defined, while
\(
\define(\{a\leftarrow{b}\},\define(\{a\leftarrow\naf{c}\},(\emptyset,\{b,c\},\emptyset,\emptyset)))
\)
is not, and the second \define\ operation is thus vacuous.

Operator \external\ takes a ground atom $a$ and transforms a system state as follows
\[
\external: \langle a,(R,I_1,i,j)\rangle \mapsto (R,I_2,i,j)
\]
where
\begin{itemize}
\item $I_2=I_1 \cup (\{a\} \setminus \Head{R})$
\end{itemize}
The \external\ operator allows for introducing atoms whose truth value can be controlled externally (via \assert, \open, and \retract)
or defined by rules later on.
Note that this is only possible if an atom is not already defined.

Let us reconsider the above example in conjunction with \external\ operations.
At first, consider
\(
\define(\{a\leftarrow\naf{b}\},\external(b,(\emptyset,\emptyset,\emptyset,\emptyset)))
\).
The state and added program induce now atom base $\{a,b\}$.
With it, the resulting state is captured by the program
\(
\{a\leftarrow\naf{b}\}
\)
having a single stable model containing $a$.
Adding afterwards $b$ via
\[
\define(\{b\leftarrow{}\},\define(\{a\leftarrow\naf{b}\},\external(b,(\emptyset,\emptyset,\emptyset,\emptyset))))
\]
results in program
\(
\{a\leftarrow\naf{b},b\leftarrow{}\}
\),
which now has a stable model containing only $b$ (but not $a$).

Operator $\release$ takes a ground atom $a$ and transforms a system state as follows
\[
\release: \langle a,(R_1,I_1,i_1,j)\rangle \mapsto (R_2,I_2,i_2,j)
\]
where
\begin{itemize}
\item $R_2=R_1 \cup \{a \leftarrow a\}$ if $a\in I_1$, and $R_2=R_1$ otherwise
\item $I_2=I_1 \setminus \{a\}$
\item $i_2^v=i_1^v\setminus\{a\}$ for 
      $v\in\{t,f,u\}$
\end{itemize}
The purpose of adding rule $a \leftarrow a$ is two-fold.
First, it prevents new rules defining $a$ to be added and,
second, assures that $a$ remains false.%
\footnote{In terms of module theory, this amounts to adding $a$ to the output atoms of the module capturing the system state.}
To see this,
observe that the compositionality criterion of \define\ prohibits the re-definition of an atom,
and that atoms must be among the input atoms to be manipulated by \assert, \open, or \retract.

As before, we observe some interrelationships between the above operations.
For a system state $S=(\RRG_1,I_1,i,j)$ and an atom $a$:
\begin{align*}
  \release(a,S) &= S\text{ if }a\notin I_1\\
  \external(a,S) &= S\text{ if }a\in I_1\cup\head{\RRG_1}\\
  \external(a,\release(a,S)) &= \release(a,S)\text{ if }a\in I_1\cup\head{\RRG_1}\\
  \define(\RRG,\release(a,S)) &= \release(a,S)\text{ if }a\in \head{\RRG}\cap(I_1\cup\head{\RRG_1})\\
  o(a,\release(a,S)) &= \release(a,S)\text{ for }o\in\{\assert,\open,\retract\}
\end{align*}


\subsection{Ground Queries}\label{sec:grdqueries}

In analogy to database systems,
we map query answering to testing membership in a (single) set.
This set is obtained by consolidating a program's stable models according to two criteria,
namely a model \emph{filter} and an entailment \emph{mode}.
A filter maps a collection of sets of ground atoms to a subset of the collection.
Examples include the identity function and functions selecting optimal models according to some objective function.
A mode maps a collection of sets of ground atoms to a subset of the union of the collection.
Although we confine ourselves to intersection and union, many other modes are possible,
like the intersection of a specific number of models or the complement of the union, etc.
Combining a filter with a mode lets us turn the semantics of a logic program, given by its set of stable models,
into a single set of atoms of interest,
for instance, the atoms true in all optimal stable models (relative to \texttt{\#minimize} statements).

The operator \query\ maps 
an atomic query $q\in\mathcal{A}$, 
an entailment mode \etmcuap, 
a model filter \filter, 
and a system state $S=(\RRG,I,i,j)$
to the set $\{\mathit{yes},\mathit{no}\}$:
\begin{align*}
  \query(q,(\etmcuap,\filter),S) & = 
    \begin{cases}
      \mathit{yes} & \text{if }\; q \in \operatorname{\etmcuap}\circ\filter(\AS(\PRG(S)))\\
      \mathit{no}  & \text{otherwise}
    \end{cases}
\end{align*}
Taking $\etmcuap=\cap$ and $\filter=\mathit{id}$ (the identity function) amounts to testing whether $a$ is a skeptical consequence of program $\PRG(S)$;
accordingly, taking $\cup$ instead of $\cap$ checks for credulous consequence.

A \emph{Boolean query} $\phi$ is an expression over $\mathcal{A}$ (in negation normal form) obtained from connectives \naf{}, $\wedge$, and $\vee$ in the standard way.
To handle such queries,
we define
a function $\Q$ as follows:
\begin{align*}
  \Q(\phi) &= \{q_{\phi} \leftarrow  \phi\} \text{ if } \phi \in \mathcal{A}\\ 
  \Q(\naf\phi) &= \{q_{\naff\phi} \leftarrow  \naf q_{\phi}\} \cup \Q(\phi)\\ 
  \Q(\phi_1\wedge\phi_2) &= \{q_{\phi_1\wedge\phi_2} \leftarrow  q_{\phi_1}, q_{\phi_2}\} \cup \Q(\phi_1) \cup \Q(\phi_2)\\
  \Q(\phi_1\vee\phi_2) &= \left\{
                               \begin{array}{l}
                                 q_{\phi_1\vee\phi_2} \leftarrow q_{\phi_1}\\
                                 q_{\phi_1\vee\phi_2} \leftarrow q_{\phi_2}
                               \end{array}
  \right\}
  \cup \Q(\phi_1) \cup \Q(\phi_2)
\end{align*}
where 
%
$\head{\Q(\phi)} \cap (\atom{R} \cup I \cup \atom{\phi})= \emptyset$
for a system state $(\RRG,I,i,j)$.

With this, the above \query\ operation can be extended to Boolean queries as follows:
\begin{equation}\nonumber
\query(\phi, (\etmcuap, \filter), S)\quad\text{ iff }\quad\query(q_\phi, (\etmcuap, \filter), \define(\Q(\phi),S)) 
\end{equation}
%
A more sophisticated way of applying the \query\ operator to a Boolean query $\phi$
in a system state $S=(R,I,i,j)$
is to execute the following sequence of operations: 
\begin{align*}
  S_1&=\external(e,S)\\ 
  S_2&=\define(\mathit{ext}(\Q(\phi),e),S_1)\\ 
  S_3&=\assert(e,S_2)\\
  v  &=\query(q_\phi,(\etmcuap, \filter),S_3)\\ 
  S' &=\release(e,S_3) 
\end{align*}
where
$e\in\mathcal{A}\setminus (\atom{R\cup\Q(\phi)} \cup I)$ is a fresh atom 
and
\(
\mathit{ext}(\Q(\phi),e) = \{\head{r} \leftarrow \body{r} \cup \{e\} \mid r \in \Q(\phi)\}
\).
As with $\query(q_\phi, (\etmcuap, \filter), \define(\Q(\phi),S))$,
we have that $\query(q_\phi,(\etmcuap, \filter),S_3)$
yields the value~$v$ of
$\query(\phi, (\etmcuap, \filter), S)$.
The respective system state~$S_3$, however, differs from $\define(\Q(\phi),S)$:
Rules from $\Q(\phi)$ are in $\mathit{ext}(\Q(\phi),e)$
annotated with a dedicated input atom~$e$.
Hence, releasing $e$ leads to a state~$S'$ such that rules
as well as head atoms of $\mathit{ext}(\Q(\phi),e)$
are obsolete and can be deleted (cf.~\cite{gekakaosscth08a}).
%
%
%

\subsection{Non-Ground Operations}\label{sec:non-ground}

So far, we considered system states and operations exclusively in the ground case. 
Let us now lay out how this lifts to the non-ground case.

As usual, a program with variables is regarded as an abbreviation for its ground instantiation.
Accordingly,
the program \RRG\ in a state $(\RRG,I,i,j)$ can be non-ground and its instantiation is left to \clingo~4 (or \gringo~4, to be precise).
Likewise, the state-changing operations \define\ and \external\ accept respective non-ground expressions. 
However, here safety of the defined rules and external declarations is an additional requirement (cf.~\cite{PotasscoUserGuide}). 
All other operations still deal with ground atoms.

Analogously, we provide means for asking non-ground queries. 
To this end, 
we extend the \query\ operator to support non-ground conjunctive queries
(but refrain from general Boolean queries for the sake of using standard safety conditions).
Specifically, 
a conjunctive query $\phi$ is a conjunction of (possibly non-ground) literals, which
can be represented by a rule
\(
q_{\phi} \leftarrow \phi
\),
provided that variable occurrences in~$\phi$ are safe.
%
With this, we can extend the \query\ operation to non-ground conjunctive queries as follows:
\begin{align*}
  \query(\phi, (\etmcuap, \filter), S) \quad\text{iff} \quad\query(q_{\phi}, (\operatorname{\etmcuap},\filter),\define(\{q_\phi \leftarrow \phi\},S))
\end{align*}
Note that this definition treats variables in~$\phi$ existentially,
and different instances of
\(
q_{\phi} \leftarrow \phi
\)
provide alternatives for deriving the target~$q_{\phi}$
of an associated atomic query.


%% file: system.tex
\section{The \aspic\ System}\label{sec:system}


Following our initial motivation to design a human-oriented
interface, we implemented the foregoing concepts and primitives in form
of an interactive ASP shell, dubbed~\mbox{\aspic}.%
\footnote{Available at \texttt{\url{http://potassco.sourceforge.net/labs.html}}.}
With this system, users can continuously enter state-changing operations to
dynamically load, define, and change an    ASP knowledge base as well as 
pose queries to explore the induced stable models.
Technically, \aspic\ is implemented in Python using 
\clingo~4\ as back-end through its Python API.
Subsequently, we illustrate the functionality and typical      workflow of \aspic\ based on an encoding for graph
coloring. 
After that, we 
give a more detailed account of the system implementation.

Initially, Listing~\ref{lst:coloring} provides an ASP encoding for $n$-coloring. 
The encoding expects a graph, that is, a set of facts over predicate \lstinline{edge}/2 representing its edges.
Given such   facts, the nodes are extracted in Line~7--8.
Then, Line~11 makes sure that each node is marked with exactly one of the $n$ colors, and Line 12 forbids that adjacent
nodes have the same color. 
Finally, the resulting stable models are projected to predicate \lstinline{mark}/2 via Line~18.
However, other than in classical one-shot solving, the encoding in Listing~1 is     not only    combinable with a fixed
graph,         but also allows an \aspic\ user to dynamically provide and change graphs.  
To achieve this, we consider instances of \lstinline{edge}/2 as input atoms, 
which is reflected by declaring them as external in Line~15.
\lstinputlisting[%
float=t,frame=single,basicstyle=\ttfamily\footnotesize,
caption={ASP encoding of $n$-coloring (\lstinline{ncoloring.lp})},%
label=lst:coloring]%
{listings/ncoloring.lp}
\lstset{keywords=[3]{dimension,horizon,t,m}}

To use \aspic\ on Listing~\ref{lst:coloring}, we run `\lstinline{aspic}' from the terminal, upon which it greets us with
its prompt prefix `\lstinline{?-}', waiting for input.
Now, we enter `\lstinline{?- load ncoloring.lp}' to load the file \lstinline{ncoloring.lp}. 
Alternatively, we could have loaded \lstinline{ncoloring.lp} immediately at invocation time by starting \aspic\ with
`\lstinline{aspic ncoloring.lp}'.
An 
overview of all supported 
commands can be obtained by entering `\lstinline{?- help}'.  
In the following, we illustrate a typical \aspic\    session, i.e., a sequence of 
\aspic\ command
interactions, documented throughout Listing~\ref{lst:graph}--\ref{lst:query4}.
\lstinputlisting[%
float=t,frame=single,basicstyle=\ttfamily\footnotesize,
caption={Building up an  initial graph by asserting     edges},%
label=lst:graph]%
{listings/graph.aspic}
\lstinputlisting[%
float=t,frame=single,basicstyle=\ttfamily\footnotesize,firstnumber=5,lastline=29,
caption={Querying with different entailment modes},%
label=lst:query1]%
{listings/query1.aspic}
\lstinputlisting[%
float=t,frame=single,basicstyle=\ttfamily\footnotesize,firstnumber=34,
caption={Interplay of a Boolean query with \assert, \open, and \retract},%
label=lst:query2]%
{listings/query2.aspic}
\lstinputlisting[%
float=t,frame=single,basicstyle=\ttfamily\footnotesize,firstnumber=62,
caption={Querying under an assumption},%
label=lst:query3]%
{listings/query3.aspic}
\lstinputlisting[%
float=t,frame=single,basicstyle=\ttfamily\footnotesize,firstnumber=75,
caption={Performing non-ground operations},%
label=lst:query4]%
{listings/query5.aspic}

First, we add an initial graph to the system by asserting the instances of \lstinline{edge}/2 given in Listing~\ref{lst:graph}.
%
Then, we ask an atomic query over the atom \lstinline{mark(1,1)},
leading to the stable model displayed in Listing~\ref{lst:query1}. 
By default, the \lstinline{query} command returns a single model matching its query along
with a corresponding satisfiability status, 
as shown in Line~6--7. 
Specifically, for an atomic query $q$ in the current system state $S$, the model returned
by \lstinline{query} is the first one found by \clingo~4 
run on the program~$\PRG(S)$,
using $q$ as an assumption (cf.~\cite{gekakaosscth08a}); 
in case of a Boolean query~$\phi$, the program and the assumption for solving change
to $\PRG(S) \cup \Q(\phi)$ and $q_\phi$. 
In either case, the satisfiability status, \lstinline{SAT} or \lstinline{UNSAT},
indicates whether some stable model matching the assumption could be found.
%
The number of models returned by \lstinline{query} can be 
changed via the \lstinline{option} command%
\footnote{Keywords are preliminary and may be subject to change in future releases.} in Line~9, 
taking \clingo~4
command line options as argument and forwarding them unaltered to the underlying \clingo~4 process.
 Here, argument `\lstinline{-n 0}' 
instructs \clingo~4 to 
enumerate all models rather than stopping at the first one. 
With that, querying again for \lstinline{mark(1,1)} in Line~11 returns the          models listed in Line~12--17.
Similarly, the entailment modes $\cup$ and $\cap$ 
can be activated via the \lstinline{option} commands in Line~20
and~27,
where the arguments `\lstinline{-e brave}' and `\lstinline{-e cautious}'
instruct \clingo~4 to return the union or
intersection of all stable models 
matching a query.
Regarding stable models including \lstinline{mark(1,1)}, the
union and intersection are obtained as results in Line~23--24 or Line~30, respectively.
To switch the 
behavior of \clingo~4 back to model enumeration, 
we then use `\lstinline{option -e auto}' in Line~33.
%
Note that, in the current \aspic\ version,
only identity is supported as model filter and, hence, always used.%
\footnote{The addition of optimization as filter is planned for an upcoming release.}

As next step, we enter the Boolean query displayed in Listing~\ref{lst:query2},
asking for stable models such that node~1 is marked with color~1 and node~3 
with color~2 or node~4 with another color than~2.
Note that the connectives $\naf{}$, $\wedge$, and $\vee$
are represented by `\lstinline{not}', `\lstinline{&}', and `\lstinline{|}'
and that subexpressions are parenthesized by square brackets.
The resulting models are shown in Line~35--37, followed by the 
corresponding satisfiability status. 
Asserting an additional edge between node~2 and~4 in Line~40 restricts outcomes
to the single model in Line~43.
On the other hand,
opening \lstinline{edge(2,4)} in Line~46 
reintroduces stable models obtained before (found in different order due to inherent
non-determinisms in search),
and also duplicates the one in which node~2 and~4 have different colors
because \lstinline{edge(2,4)} can be made true or false in this case,
while neither alternative is visible in the projected output.
After retracting \lstinline{edge(2,4)} in Line~55, the duplication
disappears, and we receive the same (number of) stable models as in the beginning.
Again note that the order of models can vary,
given that dynamic conflict information and search heuristics depend
on previous runs in multi-shot solving.

%
Listing~\ref{lst:query3} illustrates how assumptions, essentially corresponding
to atomic queries, can be incorporated.
The effect of assuming `\lstinline{not mark(2,3)}' in Line~62 is that
stable models such that node~2 is marked with color~3 are disregarded
until the assumption is canceled in Line~74.
As a consequence, the atomic or Boolean, respectively, queries in Line~64 and~70
yield proper subsets of the stable models that had been obtained before.

Finally, in Listing~\ref{lst:query4}, we make use of non-ground
expressions to which \define, \external, and \query\ operations are extended.
In Line~75, instances of \lstinline{elim}/2 are declared as external, where
the arguments~\lstinline{X} and~\lstinline{C} range over all potential nodes or
available colors, respectively.
Integrity constraints forbidding that nodes adjacent to~\lstinline{X} are marked
with~\lstinline{C} are then defined in Line~77--78.%
\footnote{Symbol `\lstinline{?}' at the end of Line~78 is an \aspic\ token indicating
the end of a collection of rules serving as argument of a \define\ operation,
so that such rules can be written in multiple lines.}
The assertions in Line~80--81 thus exclude that neighbors of node~2 and~4 are
marked with color~3 or~2, respectively.
As indicated by the outcome \lstinline{UNSAT} of asking the non-ground query in Line~83,
it is now impossible to mark node~2 or~4 with color~1.
The second non-ground query in Line~86 admits stable models
such that node~2 is marked with color~3 or node~4 with color~2,
which yields three stable models matching some instance of the query.

As 
mentioned above, 
\aspic\ is built on top of the Python API of \clingo~4,
accessible via the \lstinline{gringo} module.
Throughout a user session, the system state $(R,I,i,j)$
is captured by a 
\lstinline{gringo.Control} object. 
%
In particular,
both \lstinline{define} and \lstinline{external} commands rely on the 
API functions \lstinline{add} 
and \lstinline{ground} 
to incorporate rules and external declarations into the system state,
thus extending~$R$ or $I$, respectively.
The ``inverse'' command \lstinline{release}
uses the API function \lstinline{release_external} 
to remove an external atom from~$I$;
this includes the deletion of rules containing a released atom
in their body.
The assignment~$i$ as well as corresponding \lstinline{assert}, \lstinline{open} and \lstinline{retract} commands rely
on the API function \lstinline{assign_external} 
to manipulate (external) input atoms in~$I$.
Unlike this, assignment~$j$ is expressed by assumptions,
managed via \lstinline{assume} and \lstinline{cancel} 
and then passed to the API function \lstinline{solve},
which is invoked upon \lstinline{query} commands.
%
The \lstinline{query} command for an atomic query~$q$ is implemented
by taking the atom~$q$ as an additional (non-persistent) assumption.
%
More sophisticated Boolean or non-ground queries~$\phi$ are processed
via $\mathit{ext}(\Q(\phi),e)$ or $\mathit{ext}(\{q_\phi \leftarrow \phi\},e)$, respectively,
and the corresponding sequence of
state-changing operations given in Section~\ref{sec:grdqueries},
which are in turn mapped to API functions 
as explained above.
That is, after processing a query,     related rules             and
assumptions are withdrawn, so that the system state remains largely the
same apart from internal data structures of \clingo~4 that are updated during search.


%% file: discussion.tex
\section{Discussion}\label{sec:discussion}
We presented a simple yet effective application of multi-shot ASP solving that allows for
interactive query answering and theory exploration.
This was accomplished by means of the multi-shot solving capacities of \clingo~4.
The possibility of reusing (ground) rules as well as recorded conflict information over a sequence
of queries distinguishes multi-shot ASP solving from the basic one-shot approach.
With it, program parts can be temporarily added to the solving process until they are
withdrawn again.
A typical use case are queries that automatically vanish after having been processed.
Moreover, assertions allow 
for exploring a 
domain under user-defined hypotheses,
included in all subsequent solving processes 
until the asserted information is retracted. 

Queries and query answering play a key role in knowledge representation
and the design of intelligent agents~\cite{gelkah14a}.
Similar to our work, \sysfont{ASP-Prolog}~\cite{elposo04b} and \sysfont{XASP}~\cite{caswar01a}
provide frameworks to
interactively explore ASP programs.
In contrast to \aspic, however, both systems repeatedly invoke the stand-alone ASP solver
\sysfont{smodels}~\cite{siniso02a} to compute stable models.
While the ASP system \sysfont{dlv}~\cite{alfanipepfte11a} supports (safe) conjunctive queries
along with skeptical or credulous reasoning, it also needs to be relaunched upon each query,
whereas \aspic\ processes queries within an operative ASP solving process.
%

As future work, we plan to support non-trivial filters, e.g., optimization, and will improve \aspic's functionality based on
user feedback and experience.
In particular, we aim at generalizing the use of first-order expressions within queries
as well as state-changing operations.


%% file: acknowledgments.tex
\paragraph{Acknowledgments}

This work was partially funded 
by 
DFG
grant 
SCHA 550/9-1+2.